\documentclass[doublecol]{epl2} 
\usepackage{epstopdf}
\usepackage{subfigure}
\usepackage{graphicx}
\usepackage{amsmath}
\usepackage{xspace}
\usepackage{eqnarray}
\usepackage{bbm}

\title{Multiscale structure of the gravitational wave signal from GW150914 based on the nonextensivity $q$-triplet}
\shorttitle{Multiscale structure of the gravitational wave signal from GW150914} 

\author{C. V. da Silva\inst{1}, M. M. F. Nepomuceno\inst{2,3} and D. B. de Freitas\inst{1}}
\shortauthor{da Silva, Nepomuceno \& de Freitas}

\institute{
\inst{1} Departamento de F\'{\i}sica, Universidade Federal do Cear\'a, Caixa Postal 6030, Campus do Pici, 60455-900 Fortaleza, Cear\'a, Brasil\\
\inst{2} Departamento de Ci\^encia e Tecnologia, Universidade Federal Rural do Semi-\'Arido, Campus Cara\'ubas, Rio Grande do Norte, Brazil\\
\inst{3} Departamento de F\'{i}sica, Universidade do Estado do Rio Grande do Norte, Mossor\'o--RN, Brazil
}

\abstract{We study the first gravitational wave, GW150914, detected by advanced LIGO and constructed from the data of measurement of strain relative deformation of the fabric of spacetime. We show that the time series from the gravitational wave obeys Tsallis’s q-Gaussian distribution as a probability density and its dynamics evolve of the three associated Tsallis' indices named $q$-triplet. This fact strongly suggests that these black hole merger systems behave in a non-extensive manner. Furthermore, our results point out that the entropic indexes obtained as a function of frequency are useful statistical parameters to determine the dominant frequency when black hole coalescence is achieved. 
}

\begin{document}
\maketitle

\section{Introduction}
In September 14, 2015, Gravitational Wave Observatory by Laser Interferometer (LIGO) detected constructive interferences produced during the final fraction of a second of the merger of two black holes. Nevertheless, only on February 11, 2016, data was published for the scientific community \cite{GW,AdvancedLIGO,detector}. First predicted by Albert Einstein in 1916 as a result of the mathematical interpretation of the field  equations of general relativity, this unprecedented discovery opened a new window into the understanding of the dynamics of spacetime also called Gravitational Waves (hereafter, GW). The first gravitational wave, named by GW150914, was made from the merger of two black holes with masses of $29^{+4}_{-4}M_\odot$ and $36^{+5}_{-4}M_\odot$ observed by the two advanced LIGO \cite{AdvancedLIGO} detectors with a surprising statistical significance of 5.1$\sigma$ (further details can be found in \cite{detector}). 

The characterization of the noise in the signal is presented in \cite{noise}, and in \cite{location}, the origins of the gravitational wave are discussed. GWs that emerge from black hole mergers can be classified as evidence of either dark matter in the early universe or of continuous and stochastic sources of GWs~\cite{gw2,coyne}. These types of GW signals can be analyzed using different approaches, including Fourier transforms and the cross-correlation method \cite{coyne}.

The goal of this research is to investigate the multiscale behavior of observed strain $h(t)$ based on the analysis of Tsallis non-extensive statistical mechanics, in particular, on the estimation of Tsallis $q$-triplet, namely $\{q_{\rm stat}, q_{\rm sen}, q_{\rm rel}\}$. In the present study, we explore the behavior of the GW150914 propagating fluctuations in the spacetime curvature as a function of frequency. In addition, our investigation is based on the hypothesis of non-random patterns associated with an increase in orbital velocity towards the merger of two black holes, so does the frequency of the gravitational wave emitted. We find out that these patterns can be explained by $q$-triplet as a function of frequency \cite{deFreitas2,deFreitas}.

Our paper is summarized as follows. In Section 2, we prepare the data for two detectors from advanced LIGO. A detailed description of non-extensive frameworks and their properties are shown in Section 3. Section 4 brings the main results and discussions of the present analysis. Finally, concluding remarks are presented in the last section.

\begin{figure}
	\begin{center}
		\includegraphics[width=0.48\textwidth,trim={2.5cm 1.5cm 2.5cm 1.5cm}]{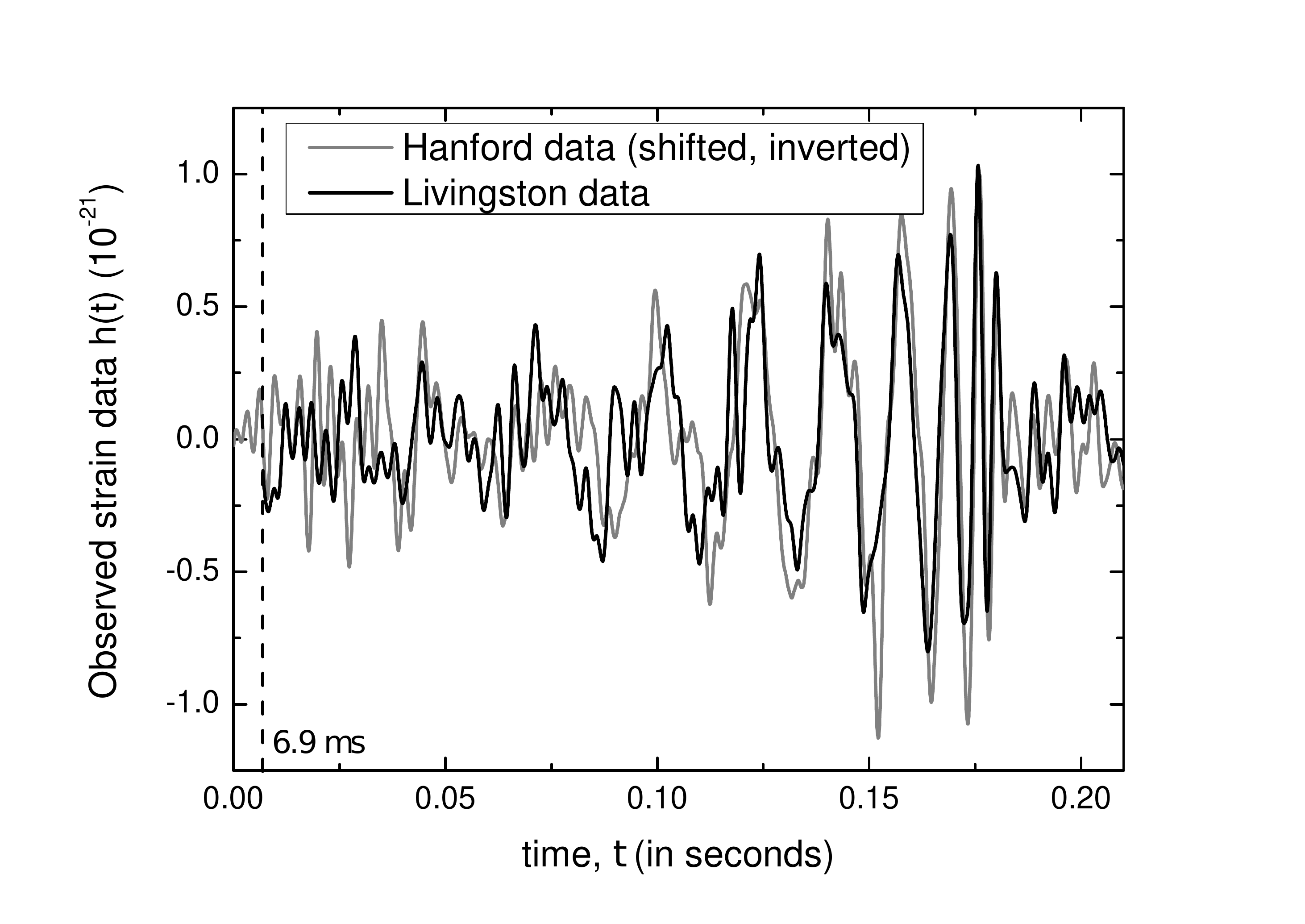}
	\end{center}
	\caption{Observed strain data in the Hanford (gray) and Livingston detectors (black). Both have been bandpass- and notch-filtered. The Hanford strain has been shifted back ($t-$6.9 ms) and inverted ($-h$). The entire visible part of the signal lasts for around 0.21 s. The dashed line indicates the shifted time of 6.9 ms.} 
	\label{fig0}
\end{figure}

\begin{figure}
	\begin{center}
		\includegraphics[width=0.38\textwidth,trim={2.5cm 1.5cm 2.5cm 1.5cm}]{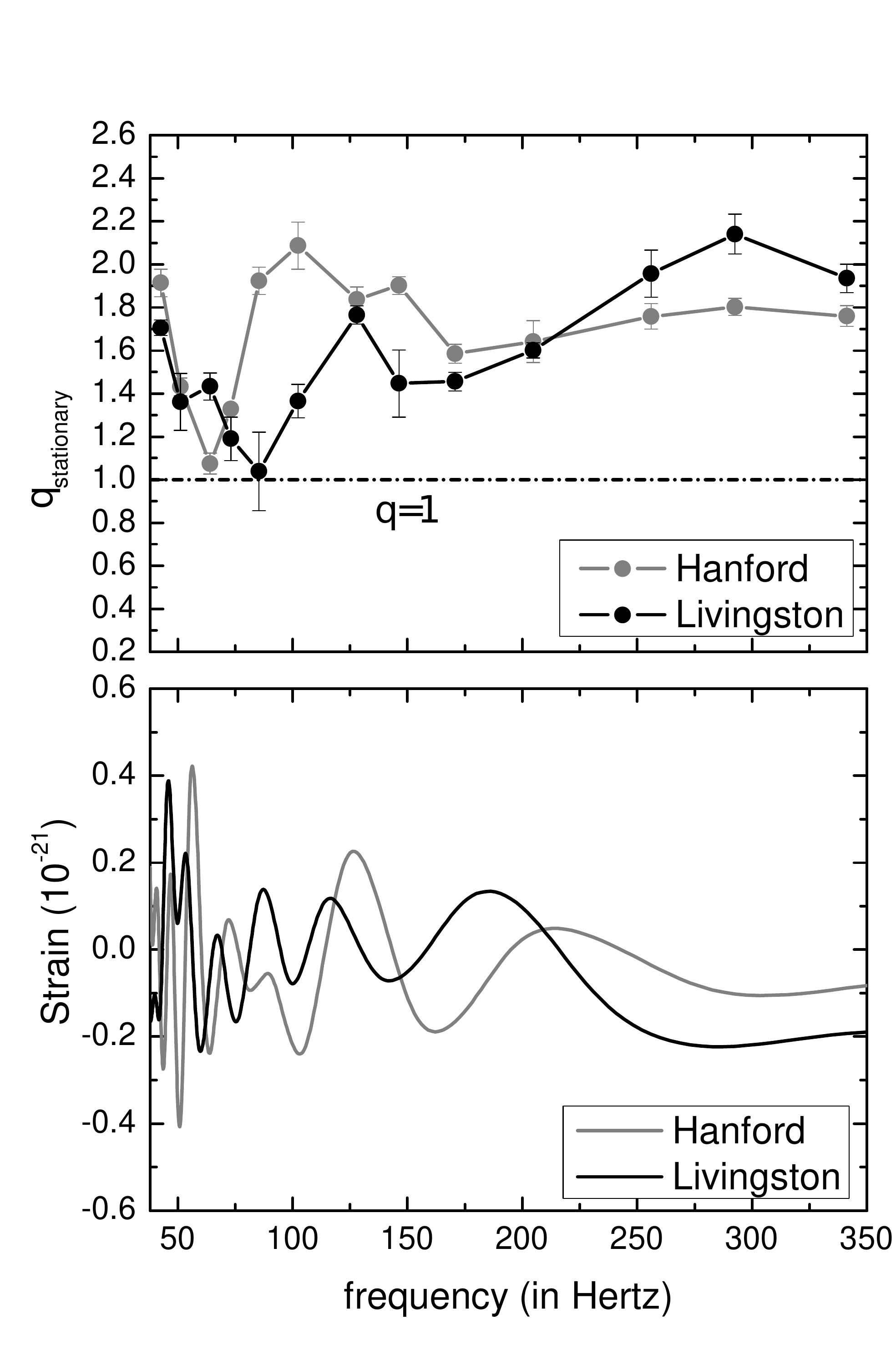}
	\end{center}
\caption{\textit{Top panel}: Evolution of the entropic index $q_{stat}$ as a function of frequency for two detectors, derived from the $q$-Gaussian distribution. $q=1$  indicates the standard Gaussian distribution. \textit{Bottom panel}: GW strain ($h$) as a function of frequency.}
\label{fig1}
\end{figure}

\section{Preparing the observed data for analysis}
The GW150914 signal is composed of several cycles of a noisy semi-sinusoidal-like time series as shown in Fig.~\ref{fig0}. In this figure, we can observe the signals of Hanford and Livingston detectors with a sampling frequency of 4096 Hz after whitening \cite{GW}. A priori, these signals present a unique chirp from 35 to 250 Hz with a duration of 0.2 s. As we can verify the wave amplitude is initially increasing, starting from around the time mark 0.30 s and, after a time around 0.42 s, the amplitude drops rapidly, and the frequency appears to stabilize. In the first region, the gravitational wave period is decreasing and, consequently, the frequency is increasing. After accounting for a 6.9 ms time-of-flight-delay in the Hanford (Washington) detector (located 3000 km from Livingston interferometer, Louisiana), the last clearly visible cycles indicate that the final instantaneous frequency is above 200 Hz as seen in Figure \ref{fig0}.

Since frequency is the most suitable parameter to study the behavior of the gravitational wave signal, we have adopted to invert the time axis (i.e., $1/t$ from Fig.~\ref{fig0}) and redraw the figure as shown in the bottom panel of Figure \ref{fig1}. After, we selected the interval of frequency from 38 to 350 Hz as explained below. The great advantage of investigating the signal from this procedure is directly compared with the results obtained from the non-extensive analysis as a function of the frequency.

Roughly speaking, the methods used to study the gravitational waves, even after discoveries from LIGO, have the focus on the Wavelet and Bayesian approaches \cite{house}. We introduce a new approach based on the non-extensive framework, which allows us to extend the investigation of fluctuations to include the effect of nearest neighbors in agreement with the delay $\tau$. The presence of large amplitude fluctuations in strain on a wide range of timescales and many large abrupt jumps in strain are two interesting features of gravitational wave time series. 

In this context, the multiscale fluctuations in observed strain $h(t)$ can be described by the increments $\Delta h(t,\tau)=h(t+\tau)-h(t)$, where $\tau$ is the time which defines the multiscale fluctuations in strain $h$. Initially, we analyzed the multiscale fluctuations following the geometrical relation $\tau=2^{n}$, where $n$ ranges from 0 to 10. The next step is to define the range of frequency. To that end, we use the relationship: $\rm frequency=\frac{1}{{\rm cadence} \cdot\tau}$ (in Hz), where the cadence is $\sim$61$\mu$s.

For the application of the procedure described above, we use gravitational wave data collected on the Gravitational Wave Open Science Center \cite{GWOSC} (GWOSC hereafter) website. The data from both LIGO detectors consist of $32\;$s and $16\;$kH of cadence. The original signals are fully affected by low frequencies with high amplitudes. To remove low and high frequencies in that LIGO sensibility is lower, we applied a band-pass frequency with the range of $38\;- 350$ Hz. It is important to highlight that the range frequency $30\;- 38$ Hz corresponds to calibrations lines. The frequencies of the 60 Hz electric power grid and its harmonics were also removed. From this cleaning in the signal, a time window with $25\;$ms of duration that includes GW150914 was analyzed in the nonextensive formalism of {\it q}-triplet. But this window also contains noise that affects the $q$-values. To determine how noise interferes with this analysis, two only-noise time windows with the same length were submitted to the same analysis. One time window was collected before the GW150914 and the other one after. All of the data processing was made using the GWpy package from Python \cite{GWpy}.

\begin{figure}
	\begin{center}
		\includegraphics[width=0.38\textwidth,trim={2.5cm 1.5cm 2.5cm 1.5cm}]{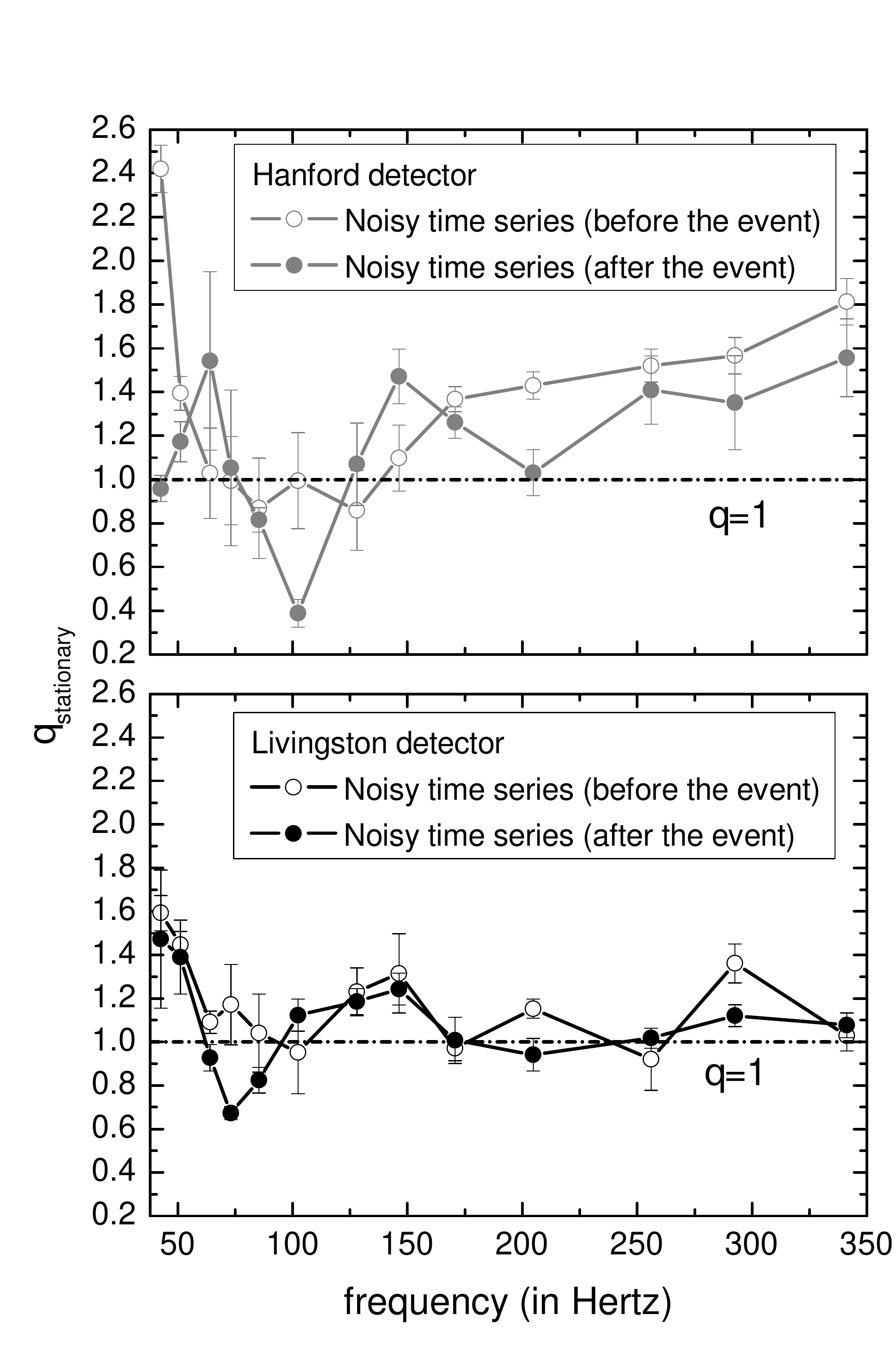}
	\end{center}
\caption{Evolution of the entropic index $q_{stat}$ as a function of frequency for two detectors segregated by pipelines before and after the event. Dash dotted line $q=1$ indicates the standard Gaussian behavior.}
\label{fig1b}
\end{figure} 

\begin{figure}
	\begin{center}
		\includegraphics[width=0.48\textwidth,trim={2.5cm 1.5cm 2.5cm 1.5cm}]{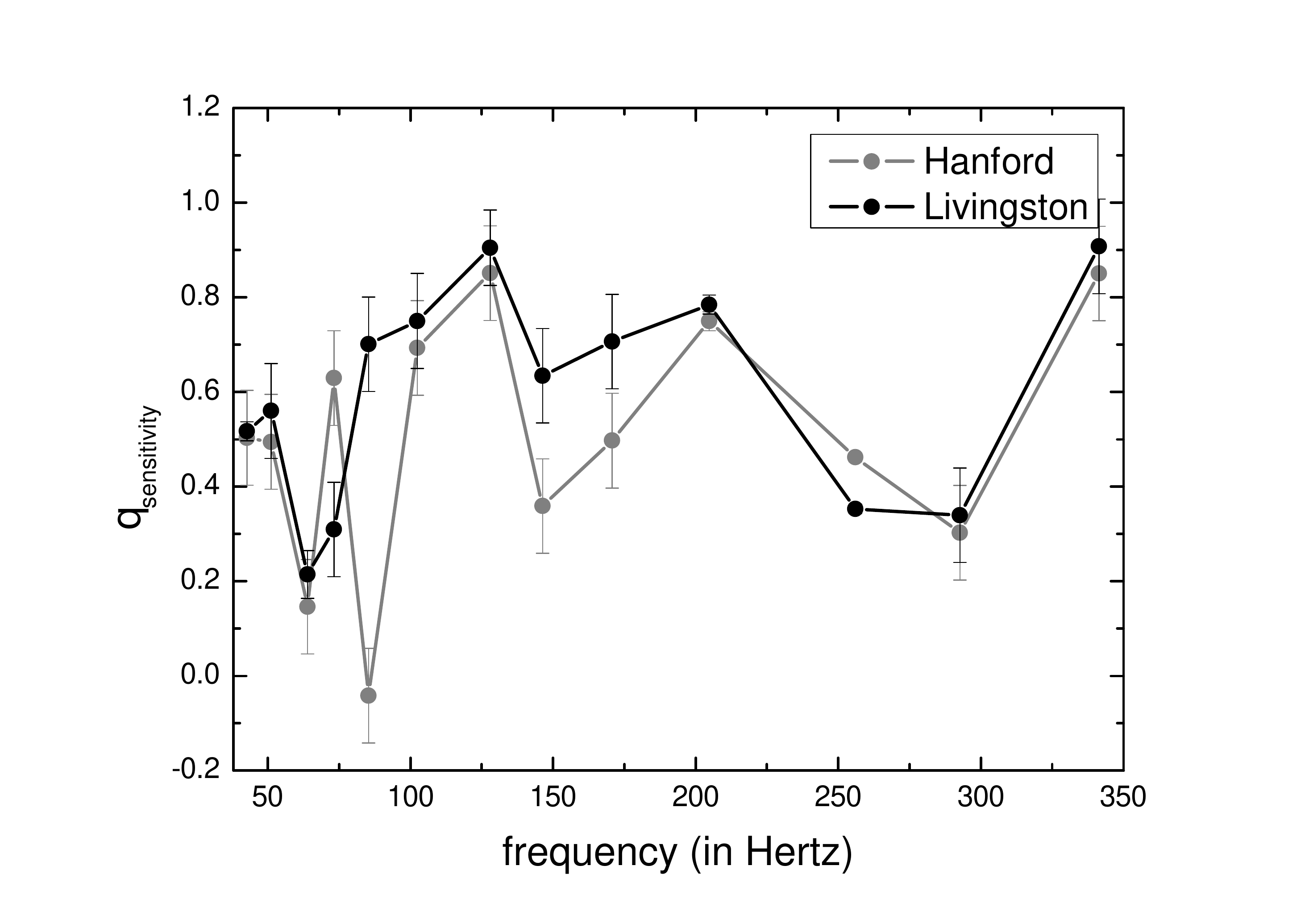}
	\end{center}
\caption{The values of $q_{sens}$ as a function of the frequency extracted from the multifractal spectrum for two detectors.}
\label{fig2b}
\end{figure}

\begin{figure}
	\begin{center}
		\includegraphics[width=0.38\textwidth,trim={2.5cm 1.5cm 2.5cm 1.5cm}]{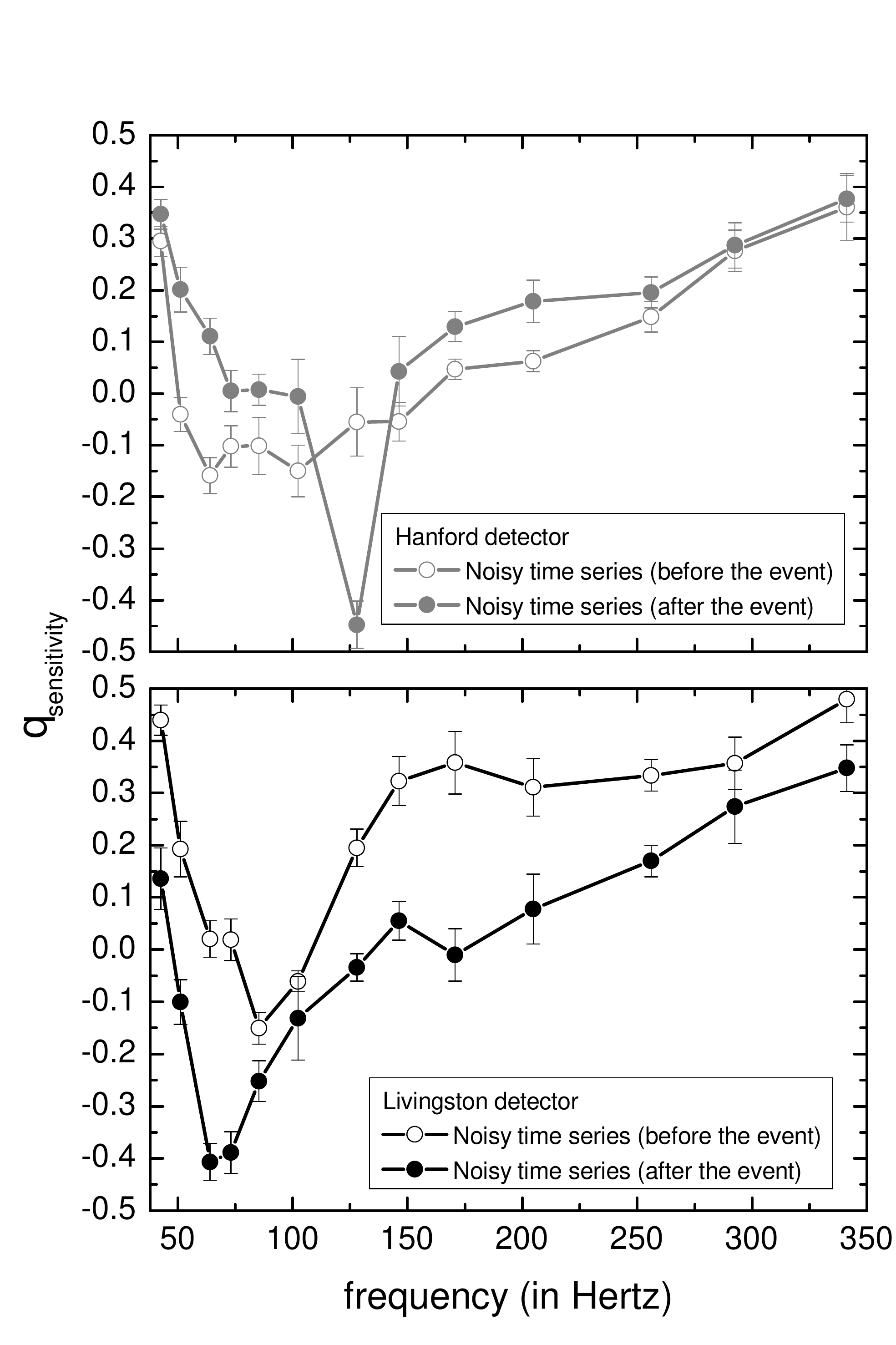}
	\end{center}
\caption{Evolution of the entropic index $q_{sen}$ as a function of frequency for two detectors segregated by pipelines before and after the event.}
\label{fig2c}
\end{figure}

\section{Nonextensive formalism}
The prototype of entropy that we are considering is the Boltzmann–Gibbs–Shannon (BGS) entropy. To mention only the most familiar statistics, this entropy has been generalized by other ``entropy-like'' indexes which emerge from approaches, such as Kolmogorov–Sinai entropy (dynamical systems)\cite{kol}, Rényi entropy (information theory)\cite{ren}, Kaniadakis entropy (relativistic kinetic theory)\cite{kani}, and Tsallis entropy (statistical physics)\cite{tsa}. Roughly speaking, the essence of these new entropic forms is to recover the BGS entropy from the Khinchin-Shannon axioms (further details see Ref.~\cite{amigo}). In particular, physical systems with long-range interactions defy the fourth axiom also called the additivity axiom, i.e., $S(A+B)=S(A)+S(B)$. To this end, a non-additive entropy uses an interaction term between the systems $A$ and $B$. In this context, the non-additivity poses the question of what mathematical properties have the ``generalized entropies'' satisfying this axiom. Our choice of Tsallis entropy form is justified by their unique properties under those axioms, in particular, the additivity axiom. As an example, the gravitational interaction is an interesting case of long-range interactions \cite{gell}.

de Freitas and de Medeiros \cite{deFreitas2}, using the same dataset of the present paper, showed that the high values of $q$ extracted from radial velocity distributions reveal effects of long-range interactions consistent with the $q$-CLT (non-extensive central limit theorem).  Because kinematics and physical properties of the space velocities of stars are defined by gravitational interaction, we chose the non-additive entropic form most appropriate and with a wide range of tested systems. As an example, Kolmogorov–Sinai,  Kaniadakis, and Rényi entropy are additive and, therefore, are at a disadvantage compared to Tsallis entropy at least for the present case.

In this context, the Tsallis' $q$-entropy, $S_{q}$, \cite{Tsallis1,Tsallis3,Tsallis4} considers that a system composed by two correlated systems $A$ and $B$ has entropy defined by
\begin{eqnarray}
\label{eq1}
S_{q}(A\oplus B)=S_{q}(A)+S_{q}(B) +(1-q)S_{q}(A)S_{q}(B),
\end{eqnarray}
where $A$ and $B$ are two independent systems and $q$ is the entropic index that characterizes generalization. When $q\rightarrow1$, the Boltzmann-Gibbs (B-G) entropy is recovered. In particular, the system in question is a tightly coupled black hole binary system and self-gravitating system.

\begin{figure}
	\begin{center}
		\includegraphics[width=0.38\textwidth,trim={2.5cm 1.5cm 2.5cm 1.5cm}]{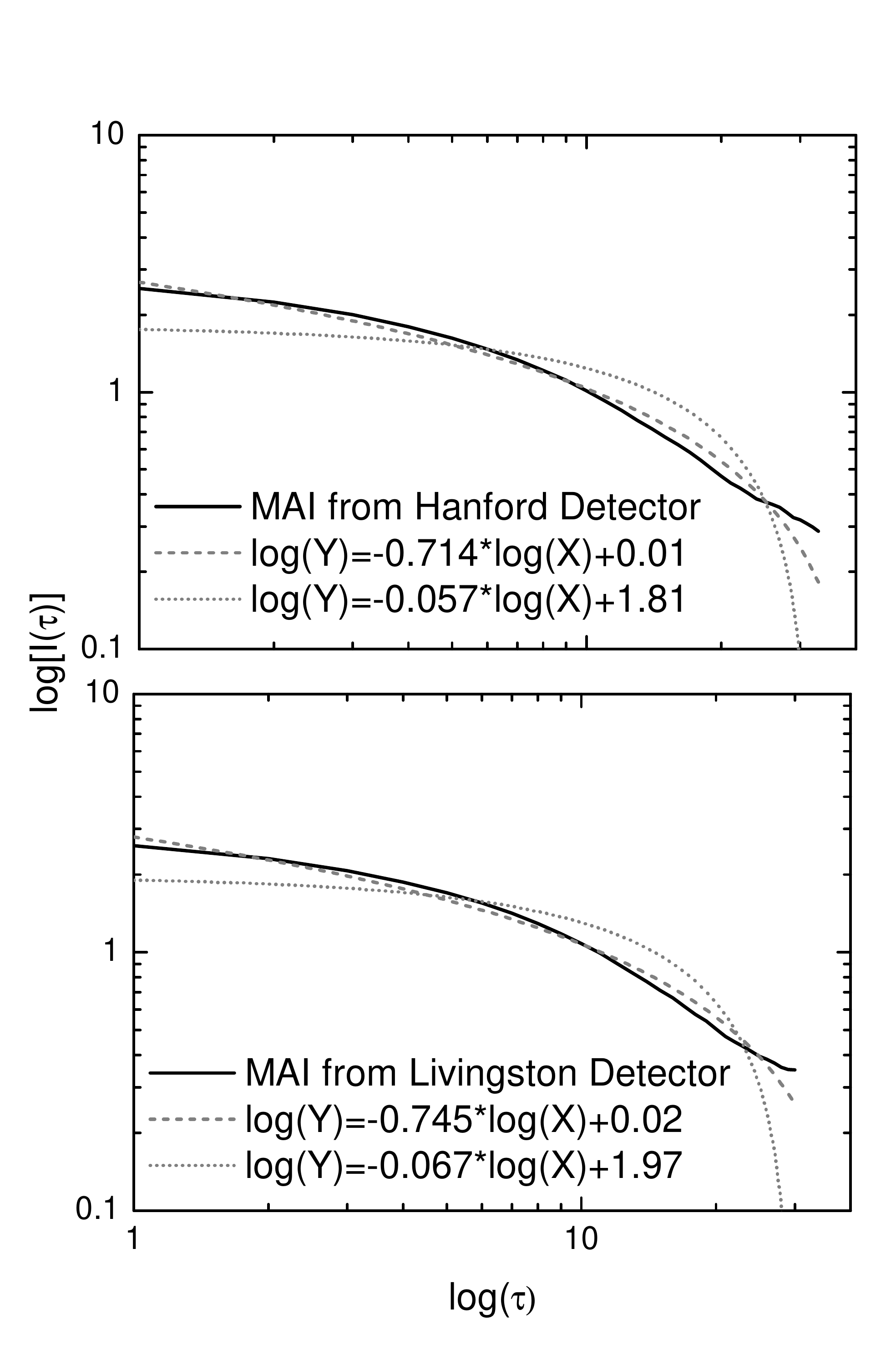}
	\end{center}
\caption{\textit{Top panel}: log-log plot of Mutual Average Information (MAI) $I(\tau)$ vs $\tau$ for the Hanford detector. \textit{Bottom panel}: Same with the previous panel but for the Livingston detector.}
\label{fig3}
\end{figure}

\subsection{Determination of the $q$--Triplet}
Several complex systems are according to the nonextensive statistical mechanics, where probability distribution function (hereafter PDF), sensitivity to the initial conditions and relaxation, namely as $q_{\rm stat}$, $q_{\rm sen}$, and $q_{\rm rel}$, are the parameters used to describe the dynamics that drive the properties of the physical system. These are referred to as the $q$-Triplet \cite{burlaga2009}.
Burlaga \& Vi{\~n}as \cite{burlaga2005} published the first paper in an astrophysical scenario of Tsallis conjecture called as a non-extensive triplet. Later, de Freitas \& De Medeiros \cite{defreitas2009} published the results of the $q$-triplet using data from the solar magnetic activity during the increasing phase of solar Cycle 23. The present study applies the same procedure used by these authors to compute the $q$--Triplet from GW150914 data.

The values of $q_{stat}$-entropic index are derived from PDF named by $q$-Gaussians, $p_{q}(x)$. These functions are obtained from the variational problem using the continuous version for the non-extensive entropy given by eq. (\ref{eq1}). $q_{\rm stat}$ is derived from PDFs by equation
\begin{eqnarray}
\label{eq2}
p_{q}(\Delta h) = A_{q}[1 +(1-q)B_{q}\Delta h^{2}]^{1/(1-q)}. 
\end{eqnarray}

For the parameter $A_{q}$, there are two conditions:

(i) $q<1$,  
\begin{eqnarray}
\label{eq3}
A_{q}=\frac{\Gamma \left[\frac{5-3q}{2-2q}\right] }{\Gamma \left[ \frac{2-q}{1-q}\right]}\sqrt{\frac{1-q}{\pi}B_{q}}
\end{eqnarray}
and (ii) $q>1$,
\begin{eqnarray}
\label{eq4}
A_{q} = \frac{\Gamma \left[\frac{1}{q-1}\right] }{\Gamma \left[ \frac{3-q}{2q-2}\right]}\sqrt{\frac{q-1}{\pi}B_{q}}.
\end{eqnarray}

The  value of $B_{q}$ is a function of variance $\sigma_{q}$ and is given by: 
\begin{eqnarray}
\label{eq5}
B_{q} = [(3-q)\sigma_{q}^{-2}]^{-1}.
\end{eqnarray}

In particular, $q_{\rm stat}$ is related to the size of the tail in the distributions \cite{burlaga2005}. This study used the Levenberg--Marquardt method \cite{l,m} to compute the PDFs with symmetric Tsallis distribution from eq. (\ref{eq2}).

The values of $q_{\rm sen}$ are obtained from the multifractal spectrum $f(\alpha)$ determined by the modified Legendre transform, through the application of the MF-DFA5 method \cite{defreitas2013q}. The parameter $q_{\rm sen}$ denotes sensitivity at initial conditions. For present purposes, we used the expression developed by Lyra \& Tsallis \cite{lyra}:
\begin{equation}
\label{1sen}
q_{\rm sen}=1-\frac{\alpha_{\rm min}\cdot \alpha_{\rm max}}{\alpha_{\rm max}-\alpha_{\rm min}},
\end{equation}
where $\alpha_{\rm min}$ and $\alpha_{\rm min}$ correspond to the zero points of the multifractal exponent spectrum $f(\alpha)$, i.e., $f(\alpha_{\rm min})=f(\alpha_{\rm max})=0$ (further details about the multifractal methods, we recommend reading the paper \cite{defreitas2017}). In the present study, we applied the same procedure used by de Freitas et al. \cite{defreitas2013q}
to compute the values of $\alpha$.

For the estimation of $q_{\rm rel}$, the autocorrelation function or the mutual average information $I(\tau)$ can be used as candidates for the observable $h(t)$. However, in contrast to the linear profile of the autocorrelation function, the mutual information includes the nonlinearity of the underlying dynamics and it is proposed as a more faithful index of the relaxation process and the estimation of the Tsallis exponent $q_{\rm rel}$. According to  the value of $q_{\rm rel}$, which describes a relaxation process, can be computed from a scale-dependent ($\tau$) mutual average information (hereafter MAI) defined by \cite{pavlos}

\begin{equation}
\label{1rel}
I(\tau)=\sum_{h_{t},h_{t+\tau}}p(h_{t},h_{t+\tau})\log_{2}\frac{p(h_{t+\tau} ,h_{t})}{p(h_{t})p(h_{t+\tau})}\sim \tau^{s},
\end{equation}
where the $q_{rel}$ index is
given by $1-\frac{1}{s}$, $s$ is the slope of the log-log plot of eq.~\ref{1rel}, and $p$ the probability distribution as before mentioned.

In summary, a nonextensive physical system is characterized by set $\left\{q_{\rm stat}; q_{\rm sen}; q_{\rm rel}\right\}\not=\left\{1; 1; 1\right\}$ where $q_{\rm stat}>1$, $q_{\rm sen}<1$, and $q_{\rm rel}>1$. For a system that obeys the B-G thermal equilibrium state, the system has the set described by triplet $\left\{q_{\rm stat}; q_{\rm sen}; q_{\rm rel}\right\}=\left\{1; 1; 1\right\}$.

\section{Results and discussions}
In this first study, we expose the scope of the methods and procedures adopted to investigate, in particular, the GW150914 gravitational wave. In a second moment, we will deal with the dozens of these GWs already available on the LIGO instrument website. In general terms, this paper analyzes step by step the behavior of the $q$-triplet that emerges from within non-extensive statistical mechanics. Its theoretical robustness allows for analyzing fine details in the geometric structure of the time series before and after the black hole coalescence process. The results that we will present below demonstrate the excellent performance of this physical theory to explain the non-linear effects that drive gravitational waves. Our analysis is an unprecedented mixing that brings together the most prosperous ``slice'' of statistical mechanics developed in recent decades.

The result of $q$-indexes as a function of frequency is presented in Figures \ref{fig1} to \ref{fig3}. These figures show the values of $q$-triplet derived from $q$-Gaussian, multifractal spectrum, and mutual average information function, respectively. 

We follow the same procedure described by Ferri et al. \cite{ferri} and found that all of the values of $q_{\rm stat}$ are between 1 and 3. Figure \ref{fig1} illustrates the behavior of the $q_{\rm stat}$–index distribution as a frequency function. It should be emphasized that this $q_{\rm stat}$ value is fully consistent with the bounds obtained from several independent studies involving the nonextensive Tsallis framework \cite{defreitas2009}. On the adopted frequency range, we can conduct a closer investigation of a possible correlation between events before, during, and after the black hole merger. With the inclusion of noisy time series before and after the event, we investigated the impact of noise on the $q_{\rm stat}$-values. For this end, we computed the $q_{\rm stat}$ for these time series as highlighted in Fig. \ref{fig1b}. At the largest frequencies, $q_{\rm stat}$ decreases and reaches a minimum value of $\sim$1 for the Livingston detector, but this behavior does not occur for the Hanford one, where a slight increase in the $q_{\rm stat}$-index seems to indicate morphological differences between the detectors. This discrepancy can be caused by the angle of incidence of the gravitational wave. The detectors are most sensitive to gravitational waves perpendicular to the plane formed by the two arms of the interferometers. As reported by Chatterjee et al. \cite{cha}, as the incidence angle becomes less than the perpendicular, the sensitivity drops \cite{cha}.  According to these authors, the curvature of the Earth causes an angle difference of around 27 degrees between the zenith direction of LIGO Hanford and Livingston which creates amplitude and phase inconsistencies. Apparently, the Hanford detector preserves the same behavior of $q_{stat}$-index when we consider the coalescence of black holes. On the other hand, this does not happen with the Livingston one.

Following the algorithm described in the paper de Freitas et al. \cite{defreitas2017}, we estimated the multifractal spectrum $f(\alpha)$ along with error bars for both GW150914 time series, and the results of $q_{\rm sen}$-index as a function of frequency are shown in Figs. \ref{fig2b}. In general, when a fractal set is homogeneous, is characterized by a single fractal dimension and a single scaling exponent. In the present study, the spectra $f(\alpha)$, calculated for GW150914 data, show a wide Holder exponent interval $\Delta\alpha$, indicating a multifractal behavior. In addition, the behavior of $q_{\rm sen}$-index for both detectors is similar with a peak at 125 Hz followed by a decrease at 150 Hz. As a result, all the values of $q_{\rm sen}$ are below $q=1$, denoting that the GW150914 gravitational wave behaves as an out-of-equilibrium thermodynamical system  (see Fig. \ref{fig2c}). At both times, before and after the event, there is a drop in the $q_{\rm sen}$index value followed by a significant increase for values greater than 100 Hz, revealing that the system tends to equilibrium as the frequency increases. We did not identify significant differences between the sensitivity to initial conditions ($q_{\rm sen}$) before and after the event, except the spurious value for the Hanford detector when black hole coalescence has already occurred. In short, this behavior may just be a reflection of what happens during the event, since the index for this detector is lower than that of Livingston. It is worth remembering that the lower frequencies are going towards the short-period fluctuations and therefore the magnitudes of the background noise order.

In Figure \ref{fig3} we present the best $\log I(\tau)$ vs $\log(\tau)$ fitting of the MAI for the GW150914 gravitational wave. With the gray dashed lines, we emphasize the power-law fitting, while with the gray dotted lines we show the exponential fitting. For a classical B-G process the MAI should decay as an exponential function. However, for two detectors, we do not find such behavior. In particular, for the Hanford detector, the MAI decays as a $q$-exponential function (power law) for lags $\tau=1-35$, as it can be seen in the top panel from Fig. \ref{fig3}. In this case, the coefficient of determination for the power law fitting was found equal to $R^{2}=0.986$, while for the exponential $R^{2}=0.801$. Thus, we use the slope $s$ of the power law fitting to estimate the $q_{\rm rel}$ index. The results showed that the $q_{\rm rel}$-index was found to be $q_{\rm rel}=1.40\pm 0.01$ (where the slope $s=-0.714\pm 0.01$), indicating a $q_{\rm rel}$-exponential decay relaxation of the system to meta-equilibrium non-extensive stationary states. Similar are the results concerning the Livingston detector (see the bottom panel from Fig. \ref{fig3}). In this case, the $q_{\rm rel}$ index was found to be $q_{\rm rel}=1.34\pm 0.02$ ($s=-0.745\pm 0.02$) with $R^{2}=0.980$ also for lags $\tau=1-35$. This result indicates a $q_{\rm rel}$-exponential decay relaxation for both detectors to meta-equilibrium non-extensive stationary states.

\section{Concluding remarks} 
In summary, we worked with gravitational wave time series for the GW150914 event for Hanford ($\rm H$) and Livingston ($\rm L$) detectors. From a fit of the Tsallis distribution to the observations with the frequency of 150 Hz we obtain the set of three indexes as listed below: 
$\left\{q_{\rm stat}; q_{\rm sen}; q_{\rm rel}\right\}_{\rm H}$=$\left\{1.90; 0.36; 1.40\right\}$ and
$\left\{q_{\rm stat}; q_{\rm sen}; q_{\rm rel}\right\}_{\rm L}$=$\left\{1.45; 0.63; 1.34\right\}$. These values clearly indicate that GW150914 gravitational wave is in an out-of-equilibrium stationary state whose physics is properly described by the $q$-statistical mechanics.

In addition, we performed a new procedure to investigate the dynamical structure of gravitational wave time series. Since the present study is limited to considering only one gravitational wave, it remains to examine some questions pointed out here with a more robust observational sample. Among these issues are some discrepancies in the values of $q_{\rm stat}$ and $q_{\rm sen}$ between the detectors, mainly before and after the event, which is necessary for a deeper investigation that can only be quantified when we test our procedure for other gravitational waves. In forthcoming communication, we will test our procedure in all GWs available on the LIGO database. Last but not least, our results suggest that the $q_{\rm stat}$ is a useful statistical parameter to determine the dominant frequency when black hole coalescence is achieved.

\acknowledgments
DBdeF acknowledges financial support 
from the Brazilian agency CNPq-PQ2 (Grant No. 305566/2021-0). Research activities of the STELLAR TEAM of the Federal University of Cear\'a are supported by continuous grants from the Brazilian agency CNPq. This research has made use of data or software obtained from the Gravitational Wave Open Science Center (gw-openscience.org), a service of LIGO Laboratory, the LIGO Scientific Collaboration, the Virgo Collaboration, and KAGRA. LIGO Laboratory and Advanced LIGO are funded by the United States National Science Foundation (NSF) as well as the Science and Technology Facilities Council (STFC) of the United Kingdom, the Max-Planck-Society (MPS), and the State of Niedersachsen/Germany for support of the construction of Advanced LIGO and construction and operation of the GEO600 detector. Additional support for Advanced LIGO was provided by the Australian Research Council. Virgo is funded, through the European Gravitational Observatory (EGO), the French Centre National de Recherche Scientifique (CNRS), the Italian Istituto Nazionale di Fisica Nucleare (INFN), and the Dutch Nikhef, with contributions by institutions from Belgium, Germany, Greece, Hungary, Ireland, Japan, Monaco, Poland, Portugal, Spain. The construction and operation of KAGRA are funded by the Ministry of Education, Culture, Sports, Science and Technology (MEXT), Japan Society for the Promotion of Science (JSPS), National Research Foundation (NRF), and Ministry of Science and ICT (MSIT) in Korea, Academia Sinica (AS) and the Ministry of Science and Technology (MoST) in Taiwan.

\end{document}